\documentclass[runningheads]{svmult}

\usepackage{makeidx}   
\usepackage{graphicx}  
\usepackage{subeqnar}  
\usepackage{multicol}  
\usepackage{physprbb}  
\makeindex             

\begin{document}
\title*{Influence of geometry in the delayed detonation model of SNIa}
%
%
%
%
\titlerunning{Influence of geometry in the delayed detonation model of SNIa}
%
\author{Domingo Garc\'\i a-Senz\inst{1,2}
\and Eduardo Bravo\inst{1,2}} 
\authorrunning{Domingo Garc\'\i a-Senz and Eduardo Bravo}
%
%
\institute{Dpt. F\'\i sica i Eng. Nuclear, Universitat Polit\`ecnica de Catalunya, Jordi Girona 1-3, M\`odul B5, 08034 Barcelona, Spain
\and Institut d'Estudis Espacials de Catalunya, Gran Capit\`a 2-4, 
08034 Barcelona}

\maketitle              

\begin{abstract}
We present several hydrodynamical simulations of thermonuclear
supernovae
dealing with multiple delayed 
detonations. The calculations were carried out in three dimensions, making 
possible to study the influence of geometry of the flame front in two  
aspects. First, the evolution of its fractal dimension during the deflagration 
phase has been followed until a critical value is reached such that the 
deflagration may turn into a detonation. Second, as the resulting detonation 
could probably be scattered through the flame, the effect of its initial 
location on the detonation propagation, final energetics and nucleosynthesis
has been explored.  

\end{abstract}

\section{Flame acceleration: from very subsonic to (maybe) supersonic
combustion rate} 

The study about how a white dwarf is disrupted by a thermonuclear
explosion, giving rise to a Type Ia Supernovae, is a hot topic in theoretical
astrophysics. Taking as starting point a carbon-oxygen white dwarf  
 near the Chandrasekhar-mass limit current 
 models must face two crucial questions: where the initial carbon runaway
 starts? 
and, once switched on, how does combustion propagate through the volume of the
white dwarf? The 
first question is related with the conditions settled in the convective core 
prior the explosion and will not be adressed here. The second question 
has deserved a lot of work during many years. In the last decade there has 
been a considerable theoretical advance owing to the increasing feasibility to
carry calculations in more
than one dimension and also because many useful ideas have been borrowed and 
adapted from the terrestrial combustion research. The central point in any 
modelization is how to accelerate the combustion from its subsonic laminar
value 
at the beginning ($\simeq 0.01 c_s$, being $c_s$\ the local sound speed) to 
a maximum value, around  $0.3c_{\mathrm s}$\ before the expansion quenches  
the nuclear energy input. It is today widely accepted that hydrodynamical instabilities
are the physical agent responsible for combustion acceleration by increasing
enormously the effective heat exchange area between fuel and ashes. 

To some extent flame propagation becomes a problem of geometry. The
ratio between the actual corrugated flame surface,$S_{\mathrm eff}$,  
and the equivalent spherical
surface encompassing the same volume can be written as: 

\begin{equation}
\frac{S_{\mathrm eff}}{4\pi r_{\mathrm b}^2}\simeq\left(\frac{v_{\mathrm max}}{v_{\mathrm l}}\right)^{(D-2){\mathrm k}}
\end{equation}

\noindent
where $r_{\mathrm b}$\ is the mean radius of the burned zone, $v_{\mathrm l}$\ is the 
microscopic laminar velocity of the flame \cite{tw92}, $\mathrm D$~ is the fractal dimension of
the effective 
surface and k is a scaling exponent which links the maximum and minimum 
scalelengths with the corresponding velocity fluctuations at these scales. For 
example, for Kolmogorov turbulence is k=3 and the minimum lengthscale 
$l_G$, called
the Gibson length in combustion theory is $l_G\simeq r_b(v_l/v_{\mathrm max})^3$. 
An interesting limiting case takes place for $D=2+1/{\mathrm k}$, then the
effective 
combustion velocity becomes independent of $v_{\mathrm l}$. This behaviour
has been observed 
in terrestrial experiments dealing with turbulent chemical flames in 
the corrugated regime where $D=7/3$. In the 
context of supernovae there are more physical mechanism affecting the flame 
surface other than turbulence but the above statement still approximately
holds \cite{k95}. 
This result opens a way to model the explosion by simply substituing the
laminar velocity $v_\mathrm l$~ by another one,$v_\mathrm b$, high  enough as to rise $l_{\mathrm 
G}$\ to the level of resolution of the hydrocode. The adoption of such {\sl  
baseline velocity} 
$v_\mathrm b$, instead of $v_\mathrm l$~ is equivalent to have a subgrid below the level of 
resolution of the code (see Sect. 2).

The above picture gives a combustion front which moves very subsonically 
at the beginning, when instabilities are not so strong, but rapidly accelerates
when instabilities plenty develop. However, in order to get the adequate 
explosion energetics and synthesize a substantial amount of intermediate-mass 
elements the effective front velocity must reach a quite high value, 
$v_{eff}\ge 10^3$\ km.s$^{-1}$\ once the density at the front location  
declines below $\rho=5~10^7$\ g.cm$^{-3}$. Multidimensional models based 
on pure deflagrations have traditionally had 
some difficulty to obtain such precise effective velocity profile, especially
at low densities
(but see J. Niemeyer this volume). One way to cure this drawback is to 
turn the deflagration into a supersonic detonation at late times, as many 
one-dimensional models have shown \cite{k91}. Even though the physical mechanism 
driving such transition has not been found yet it could be related to two 
possible causes: 1) spontaneous burning of a critical mass of carbon when 
combustion enters in the distributed regime \cite{nw97,kow97}, 2) an effective velocity  
exceeding the maximum  value for a Chapman-Jouguet (CJ) deflagration 
to be
stable \cite{nw97,bgs94}. Although it is not evident that the CJ limit, strictly defined 
for planar fronts, is extrapolable to different geometries using the
effective velocity instead the laminar one  
we have adopted this second possible mechanism 
in order set a practical criteria to select those points of the flame which
 turn into a detonation.

Up to now there are very few multidimensional calculations dealing with the 
delayed detonation scenario \cite{bgs94,l99}. Two dimensional simulations 
\cite{l99} are
enforced  
to detonate carbon in a single region located near the singularity axis. Full 
three-dimensional calculations are, however, desirable because their ability 
to handle  multipoint ignitions and  better representing the important
previous deflagration phase. When calculated in three dimensions the 
detonation trajectories can be significatively altered by the irregular 
distribution of fuel and ashes set in the deflagration phase and by
the detonation-detonation and 
detonation-deflagration wave interactions. As these effects  will affect the
outcome of 
the explosion it is worth to make a first exploration of this scenario.

\section{Method of calculation and models}

The simulations were carried out using the SPH technique adapted to handle both 
steady thermal waves (flames) and shocks (detonations)\cite{gsbs98}. We included a 
realistic physics: EOS consisting of relativistic partially degenerated       
electrons with pair corrections, ions as ideal gas plus Coulomb corrections 
and radiation. The nuclear part is a small nuclear network of 9 nuclei from 
helium to nickel, wherever the temperature exceeded five billion degrees the 
material is isochorically processed to the nuclear statistical equilibrium, 
electron captures were allowed in this regime. Detailed nucleosynthesis 
was calculated by postprocessing the output of the hydrodynamics  of each 
model.

The advance of the flame in the deflagration stage was simulated by expanding 
the flame thickness (much lesser than 1 cm!) to the actual resolution of the 
code (several dozens of km) by adequately rescaling the actual microscopic 
conductivity and nuclear energy rate in the energy equation. In addition, 
ellipsoidal kernels were used in the detonation phase in order to increase 
the resolution in this regime, typically an improvement
between 2-3 is achieved.

The initial model is an isothermal  white dwarf of $1.36$ M$_\odot$ and 
$\rho_c=1.8~10^9$\ g.cm$^{-3}$ in hydrostatic equilibrium. The first stages   
of the explosion were followed by using a one-dimensional Lagrangian code 
until the central density declined to $\rho_c=1.4~10^9$\ g.cm$^{-3}$. At 
this point the structure was mapped to a 3D distribution of 250,000 particles, 
and the velocity field around the flame contour perturbed by a sample of
20 sinusoidal functions 
of different amplitude and wavelength. Afterwards the evolution was  
followed with the SPH code. 

During the progression of the calculation the geometrical features of the 
flame front were tracked by calculating the main scaling parameters of the 
surface such as $l_{\mathrm min}, l_{\mathrm max} {\rm and}~ D$, using the method 
described in \cite{gsbs98}. The value of $l_{\mathrm min}$\ can be used to set the baseline 
velocity $v_b$. In principle an optimum value can be found by solving the
equation $l_{\mathrm min}(v_b,\vec r)= h(\vec r)$, where $h$~ is the
smoothing 
length parameter.
 Even though $v_b$\ is a time-dependent  
local quantity we have found that, on average, a constant value of
$v_b\simeq 60-70$\ km.s$^{-1}$\ is enough to move $l_{\mathrm min}$\ (or   
equivalently the Gibson length, $l_G$) to the coarse resolution of the SPH code.

\begin{table}
\caption{Main features at the end of the deflagration phase}
\begin{center}
\renewcommand{\arraystretch}{1.4}
\setlength\tabcolsep{5pt}
\begin{tabular}{llllll}
\hline\noalign{\smallskip}
t & $M_{\mathrm b}$ &
Total energy & $M_{\mathrm Ni}$ & $M_{\mathrm C+O}$ & $M_{\mathrm Si}$ \\
seconds & $\left({\mathrm M}_\odot \right)$ & $\left(10^{51} {\mathrm erg}\right)$ &
 $\left({\mathrm M}_\odot \right)$ & $\left({\mathrm M}_\odot \right)$ &
 $\left({\mathrm M}_\odot \right)$ \\
\noalign{\smallskip}
\hline
\noalign{\smallskip}
 1.54 & 0.52  & 0.18 & 0.27 & 0.65 & 0.07 \\
\hline
\end{tabular}
\end{center}
\label{tab1}
\end{table}

\subsection{Deflagration phase}

\begin{figure}
\begin{center}
{
 \centering
 \leavevmode
 \columnwidth=.45\columnwidth
 \includegraphics[width=\columnwidth]{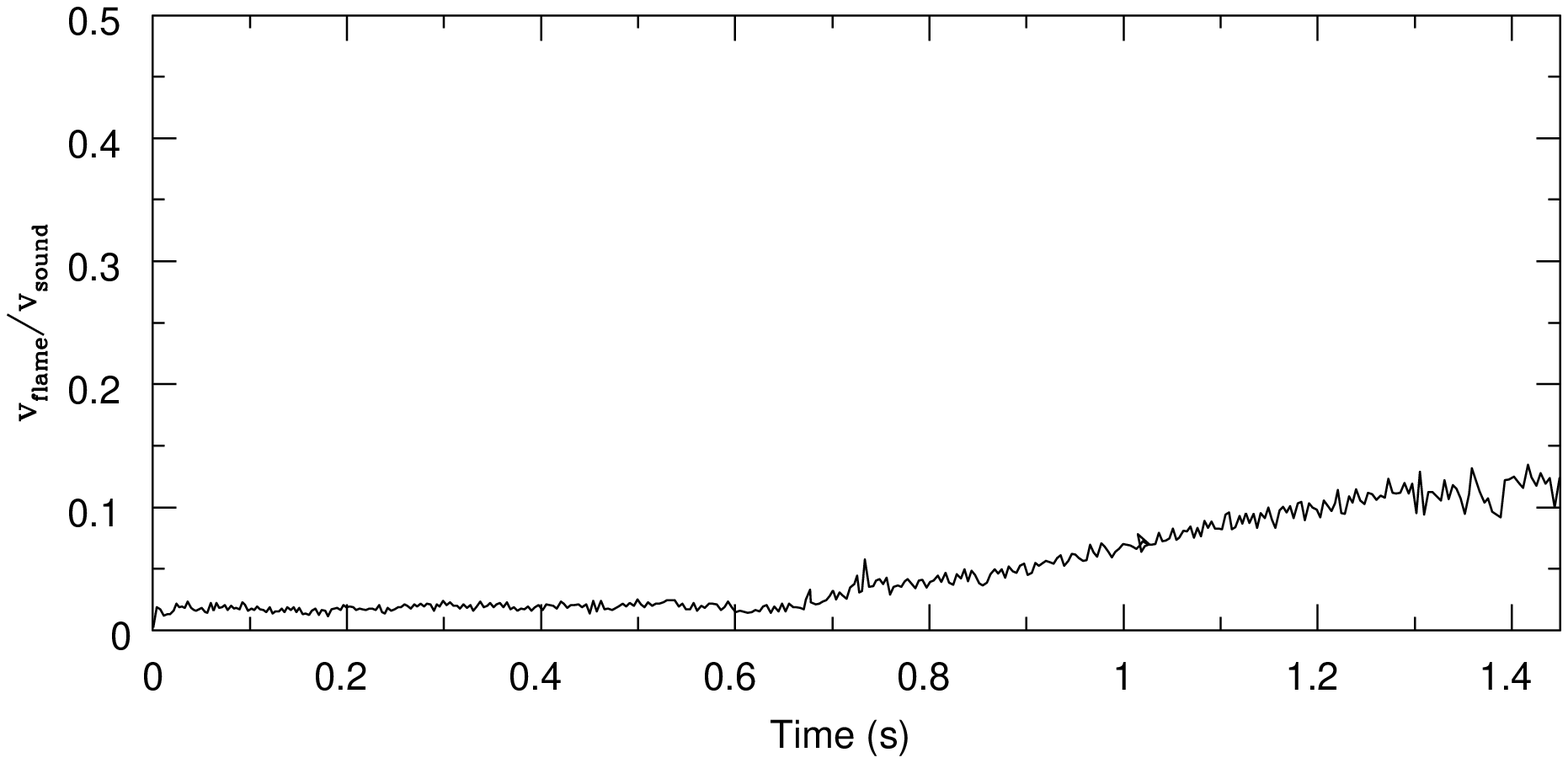}
 \hfil
 \includegraphics[width=\columnwidth]{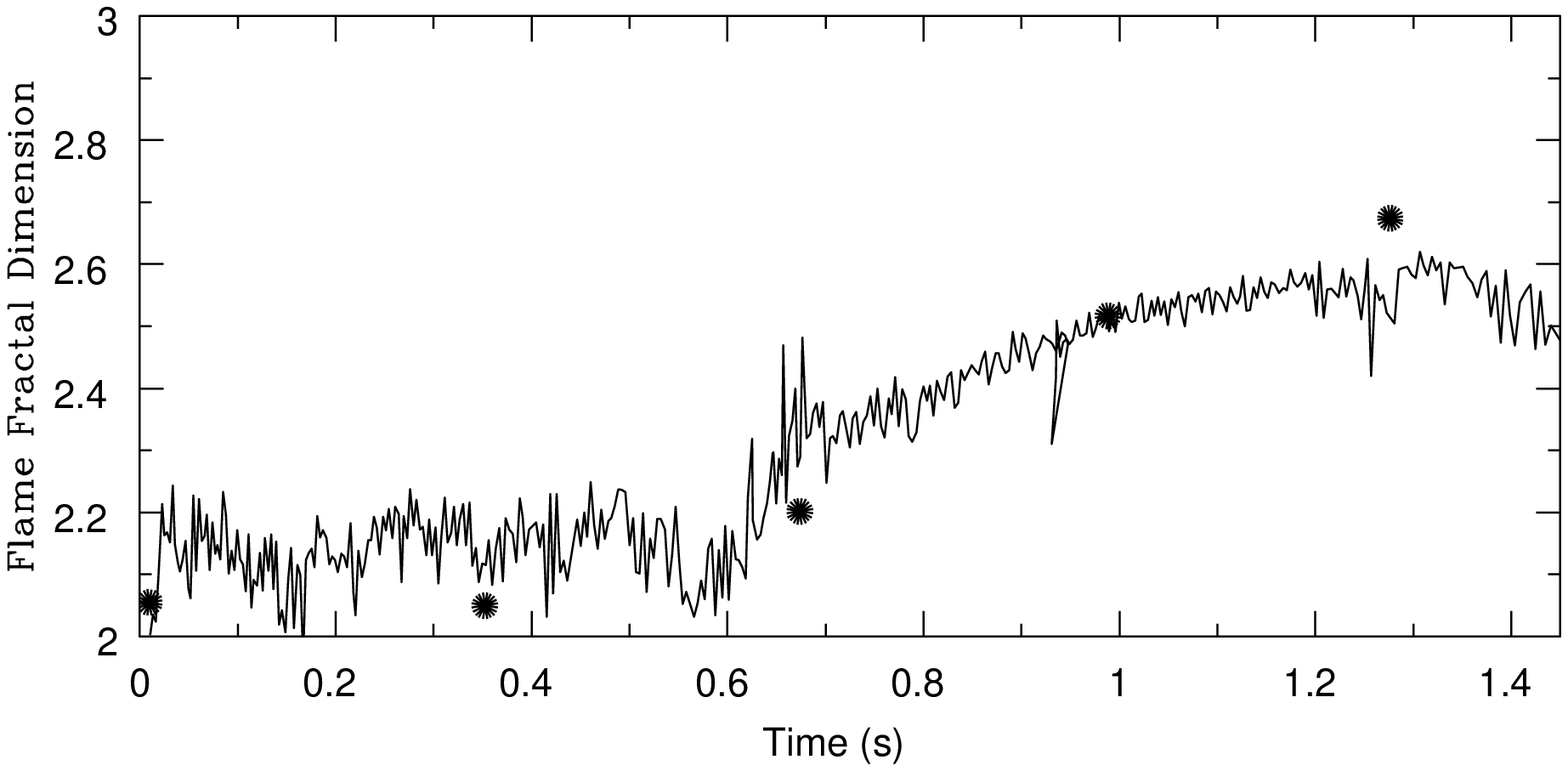}
}
\end{center}
\caption{Evolution during the deflagration phase.
(\textbf{a}) Effective velocity of the flame in units of the local sound speed. 
(\textbf{b}) Evolution of the fractal dimension of the flame in two
{\sl independent}
approximations: correlation dimension calculated as in \cite{gsbs98} (big 
dots) and derived from the burning mass rate given by the SPH 
(continuum line). 
Both calculations are consistent.
}
\label{eps1}
\end{figure}

The spherical symmetry was kept during the first half of second of the
evolution, 
the flame moving with $v_b$, almost the laminar  
value near the center. Afterwards the typical fingers of the Rayleigh-Taylor
instability in the linear regime developed. When t=0.7 s the effective velocity 
and the nuclear rate input began to increase (Figs. 1,2). After one second the
front displayed the mushroom like shape characteristic of the non-linear
regime. At 
approximately t=1.1 s the nuclear energy input reached a maximum of
$1.3~10^{51}$ ergs/s, the effective velocity of the front
being $\simeq 0.12 c_s$. A maximum value of $v_{\mathrm eff}=0.14 c_s$\ was reached at 
t=1.41 s. From here on the effective velocity declined owing to the expansion. 
The main features at the end of this phase are summarized in Table 1. 

\begin{table}
\caption{Calculated models}
\begin{center}
\renewcommand{\arraystretch}{1.4}
\setlength\tabcolsep{5pt}
\begin{tabular}{llllll}
\hline\noalign{\smallskip}
Model & Detonation criteria & 
$E_{kin}$ & $M_{\mathrm Ni}$ & $M_{\mathrm C+O}$ & $M_{\mathrm Si}$ \\
 &at $\left<\rho_{\mathrm flame}\right>=2~10^7$ g.cm$^{-3}$ & $\left(10^{51} {\mathrm erg}\right)$ &
 $\left({\mathrm M}_\odot \right)$ & $\left({\mathrm M}_\odot \right)$ &
 $\left({\mathrm M}_\odot \right)$ \\
\noalign{\smallskip}
\hline
\noalign{\smallskip}
A & D$_{\mathrm fractal}>2.5$& 0.75 & 0.54 & 0.34 & 0.16 \\
B & r$<1.3~10^8$~cm& 0.48 & 0.43 & 0.48 & 0.1 \\
C & $1.7~10^8 {\mathrm cm}<r<2.0~10^8{\mathrm cm}$ & 0.51 & 0.42 & 0.45 & 0.14 \\
D &  $2.4~10^8 {\mathrm cm}<r<2.8~10^8{\mathrm cm}$& 0.33 & 0.34 & 0.57 & 0.09 \\
\hline
\end{tabular}
\end{center}
\label{tab2}
\end{table}

The evolution of the fractal dimension is also shown in Fig. 1. After
an initial  
period where the fractal dimension of the flame surface remained very close 
to two it rapidly increased for $t>0.7$\ s. At that time the averaged density
of the flame 
dropped below $5~10^8$\ g.cm$^{-3}$. After t=1.1 s the fractal dimension rose 
slowly until a value $D\simeq 2.6$, similar to that associated with the 
RT instability in the non-linear regime, $D_{\mathrm RT}=2.5$.

\begin{figure}
\begin{center}
{
\centering
\leavevmode
\columnwidth=.45\columnwidth
\includegraphics[width=\columnwidth]{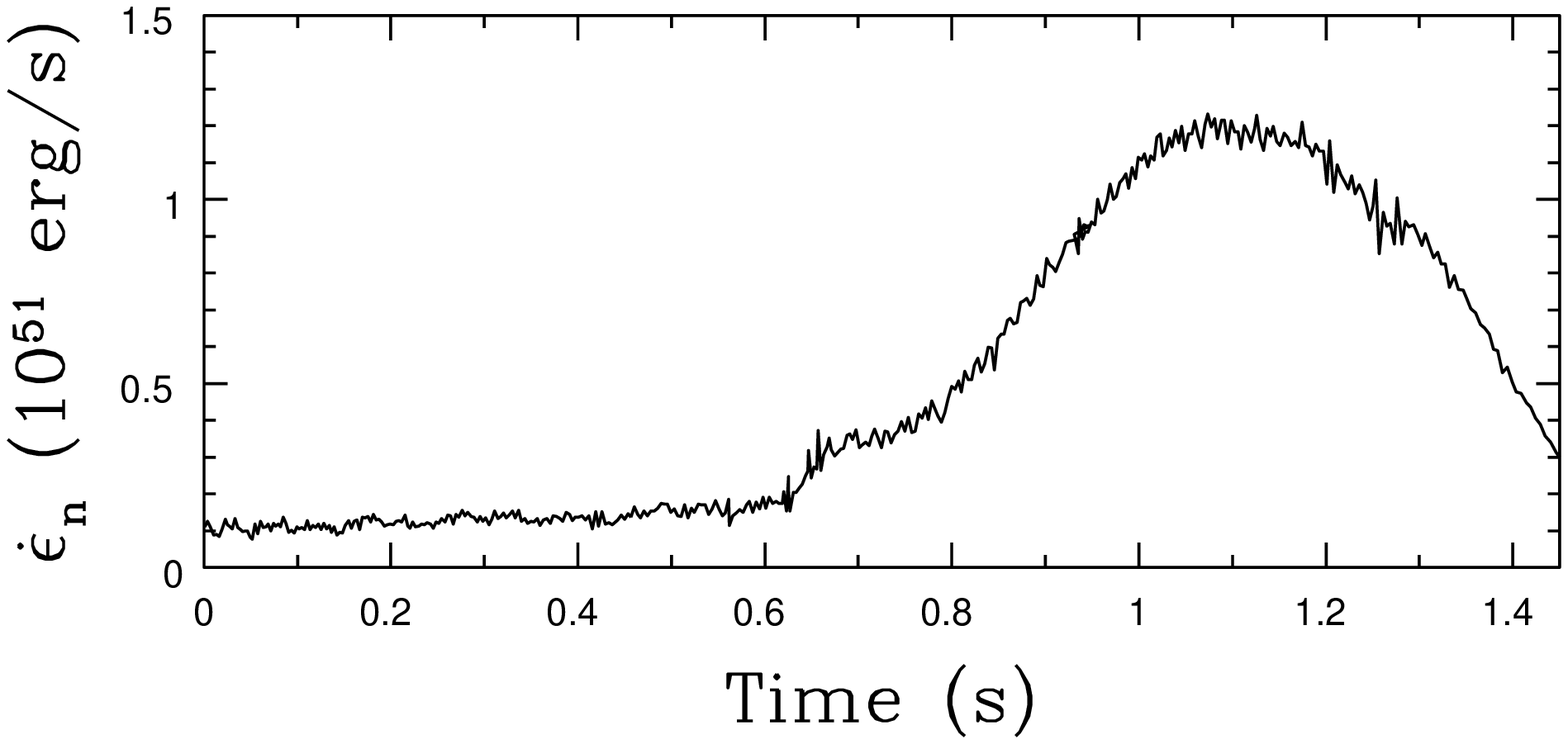}
\hfil
\includegraphics[width=\columnwidth]{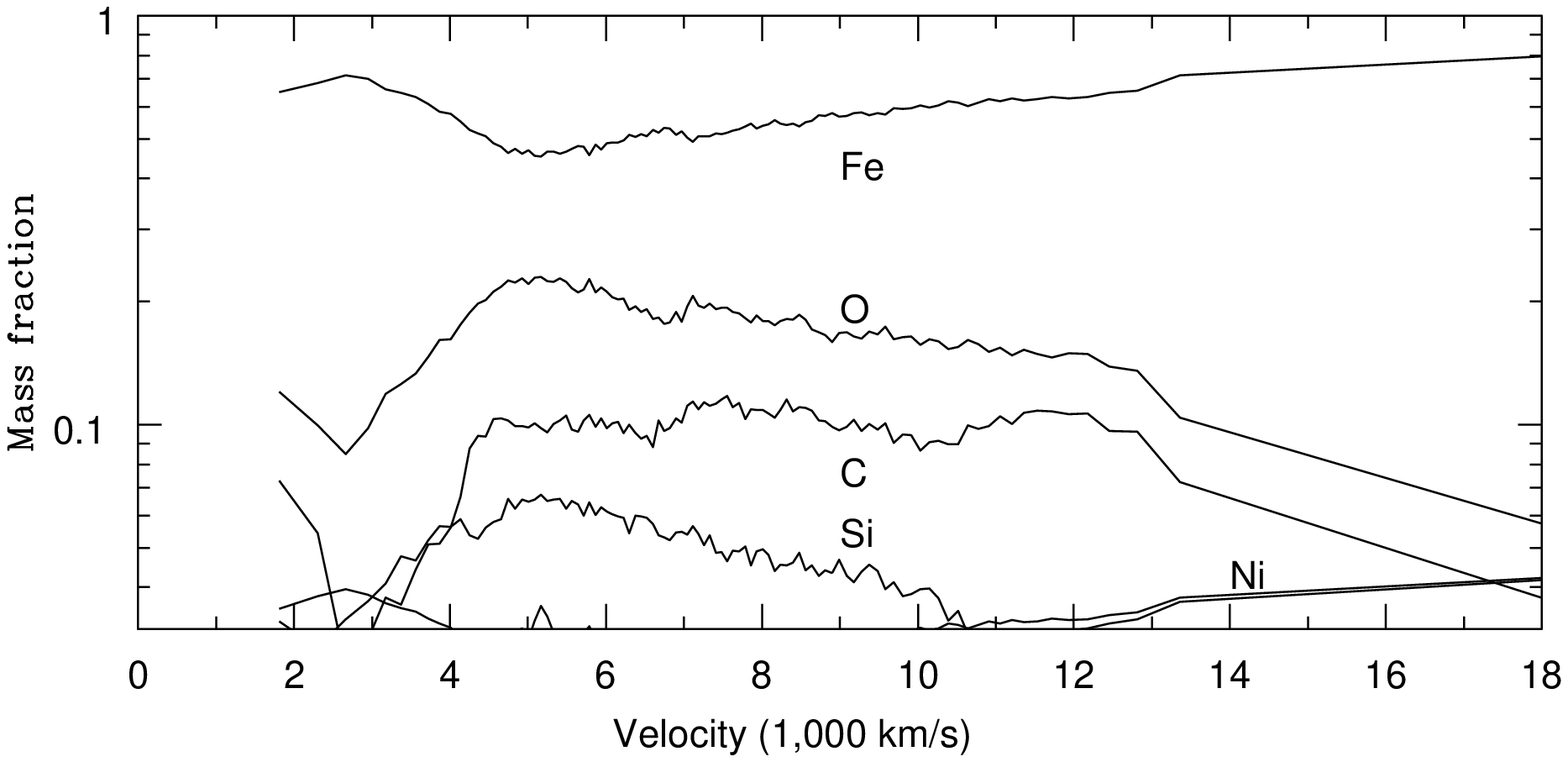}
}
\end{center}
\caption{
(\textbf{a}) Rate of released nuclear energy in the deflagration phase. 
(\textbf{b}) Distribution of $^{56}Fe (^{56}Ni)$, $^{28}Si$~and $^{12}C, 
^{16}O$~ in velocity space} 
\label{eps2}
\end{figure}

\subsection{Detonation phase}

When $<\rho_{\mathrm flame}>=2~10^7$\ g.cm$^{-3}$~the flame surface was rather  
complex, displaying many regions with high fractal dimension. In fact, a 
multifractal analysis showed that  $\simeq 20\%$~ of the mass within the 
flame had $D>2.5$. A high value of ${\mathrm D}$~ means that the local effective velocity 
of the front is higher than the average value of $v_{\mathrm eff}$~making  
possible the transition to a detonation by the second mechanism commented in 
Sect. 1. A rough estimation of the minimum fractal dimension needed  
 to exceed the Chapman-Jouguet limit was given in \cite{bgs94} where it 
was assumed that turbulence ($D_{\mathrm turb}=7/3$) dominates in those   
scalelengths not directly resolved by the hydrocode. According to that analysis,
a fractal dimension $D>2.5$~ when
 $\rho\simeq 2~10^7$~g.cm$^{-3}$~ could make  
possible the transition. We took this criteria as a procedure to pick up the
points 
of the flame prone to turn into a detonation in Model A of Table 2, our 
main calculation. In models B, C, D we explore the influence of the 
altitude in the outcome of the model by artificially detonating {\sl all} 
the mass within the flame at different radii.

\begin{figure}
\begin{center}
\includegraphics[width=.3\textwidth, angle=270]{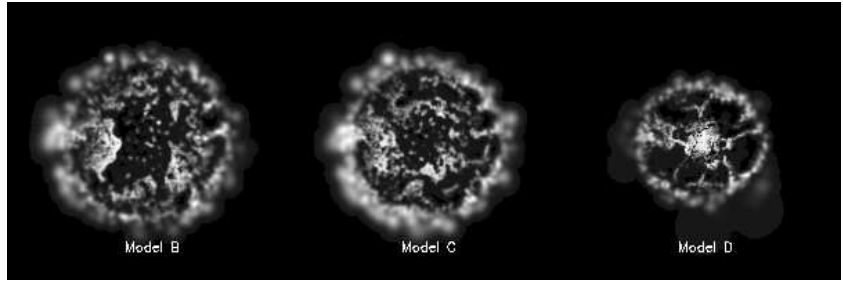}
\end{center}
\caption{Final abundance of $^{12}C+^{16}O$~for models 
B,C, and D in a meridian slice}
\end{figure}

Model A gave an acceptable SNIa model, as can be seen in Table 2 and
Fig. 2b. The resulting explosion energy was a little weak but not bad and 
the $^{56}Ni$~ amount is enough to power the light curve. The final abundance 
of intermediate-mass elements  
was right although they showed a large spread in velocity space 
as 
can be seen in Fig. 2. Nevertheless there remained $0.34$~M$_{\odot}$~of unburnt
carbon and oxygen chiefly due to both, the shielding effect of the ashes 
produced during the deflagration phase and the degradation of the resolution 
when detonation propagates in the low-density regions.
The detonation altitude largely influences the outcome of the 
explosion as shown in Table 2 and Fig. 3. Models B  and C 
were not totally unsatisfactory but Model D  
 was clearly bad, the explosion was too weak 
and an unacceptable amount of unburnt $^{12}C+^{16}O$~was left even at the 
center of the white dwarf (Fig. 3). 

\section{Conclusions}

The explosion of a Chandrasekhar-mass white dwarf has been simulated in three 
dimensions. The adopted explosion mechanism was a deflagration followed by 
multiple detonations when the mean density of the flame front declined below 
$\rho=2~10^7$~g.cm$^{-3}$. In addition the dependence of the main observables 
of the explosion against the geometrical location of the initial detonating spots 
has been explored. Our best model (model A of Table 2) is an acceptable 
model for Type Ia Supernovae albeit the explosion was a little weak. A relevant 
feature is that either $^{56}Ni$~ as well as intermediate-mass elements
showed   
a large dispersion in  velocity space. A negative point was the large amount of
 unburnt carbon and oxygen which were left in isolated pockets. The sensitivity 
 of the outcome on the initial detonation distribution is strong, as discussed 
 in Sect. 2. This fact relies in the irregular distribution of fuel and 
 ashes settled by the initial deflagration and may provide a natural way
 to explain 
 the Type Ia supernovae diversity.

There is still needed  a lot of work to confirm or reject
the delayed detonation   
scenario. In particular asynchronous multipoint detonations with more 
resolution must be attempted to shape the model. On the other
hand there are some expectatives that deflagration 
models dealing with active turbulence could also give right models for 
thermonuclear supernovae without invoking detonations 
(J.Niemeyer and Bravo and Garc\'\i a-Senz this volume). For the time being it is necessary to
investigate both mechanisms. 

This work has been benefited from the MCYT grants EPS98-1348 and AYA2000-1785 
and by the DGES grant PB98-1183-C03-02.

%


\begin{thebibliography}{8.}
\addcontentsline{toc}{section}{References}

\bibitem{tw92} F.X. Timmes and  S.E. Woosley: ApJ \textbf{396}, 649 (1992) 

\bibitem{k95} A.M. Khokhlov: ApJ \textbf{449}, 695 (1995) 

\bibitem{k91} A.M. Khokhlov: A\&A \textbf{245}, 114 (1991)

\bibitem{nw97} J. Niemeyer and  S.E. Woosley: ApJ \textbf{475}, 740 (1997)

\bibitem{kow97} A.M. Khokhlov, E.S. Oran and J.C. Wheeler: ApJ
\textbf{478}, 678 (1997) 

\bibitem{bgs94} E.Bravo, D. Garc\'\i a-Senz:'Asymmetrical delayed detonation from a 3D hydrosimulation of a white dwarf explosion'. In 
\emph {Cosmic Chemical Evolution. 
Proceedings of the 187th Symposium of the IAU, Kyoto}, ed. by 
K. Nomoto and J.W.Truran (Dordrecht: Kluwer Academic Publishers, 2002), p220

\bibitem{l99} E. Livne: ApJ \textbf{527}, L97 (1999) 

\bibitem{gsbs98} D.Garc\'\i a-Senz, E. Bravo., N. Serichol: ApJSS
\textbf{115}, 119 (1998)

\end{thebibliography}
\end{document}